\documentclass[9pt,twocolumn,twoside]{gsajnl}

\usepackage{epstopdf}
\usepackage{float}
\usepackage{textgreek}
\usepackage[section]{placeins}
\usepackage{gensymb}
\usepackage{siunitx}
\usepackage{pgfplots}
\usepackage{relsize}

\articletype{inv} 

\runningtitle{Report} 
\runningauthor{Fujiwara}

\title{A study on Non-Performing Assets Cases and Cryptocurrency in Japan}

\author[1,$\ast$]{Burina Fujiwara} 

\affil[1]{Department of Economics and Finance, College of Business, City University of Hong Kong}

\correspondingauthoraffiliation[$\ast$]{Corresponding author: Department of Economics and Finance, College of Business, City University of Hong Kong, Kowloon, Hong Kong. \href{fujiwara.b@my.cityu.edu.hk}{fujiwara.b@my.cityu.edu.hk}}

\begin{abstract}
The economic bubble bursting resulted in a large number of non-performing loans in Japanese financial institutions, which weakened their functions and prevented them from extending credit for normal economic activities. However, cryptocurrency operations are thriving in Japan. 
 
In this way, this paper focuses on non-performing assets and cryptocurrencies. The goal is to use literature analysis methods to summarise the development process, types of issuance, mechanisms, evaluation models, application scenarios, and trends in how cryptocurrencies are supervised.
\end{abstract}


\dates{\rec{25 09, 2022} \acc{25 09, 2022}}

\begin{document}
\urlstyle{same}

\maketitle

\thispagestyle{firststyle}
\vspace{-13pt}

\section{Introduction}
Japan has the world's most advanced currency-buying platform and cryptocurrency regulatory environment and recognises Bitcoin and other cryptocurrencies as legal property in accordance with the Payment Services Act (PSA) and the Financial Instruments and Exchange Act (FIEA). Under these regulations, cryptocurrency exchanges in Japan must register and adhere to standard AML/CFT (Anti-Money Laundering/Combatting the Financing of Terrorism) requirements. Given that Japan is the world's largest market for buying and selling bitcoin, the National Tax Service determined in December 2017 that Japanese cryptocurrency gains should be classified as "other income", and investors should pay tax accordingly.

The total market value of encrypted assets exceeded two trillion dollars by September 2021, a nine-fold increase since the beginning of 2020. The ecosystem is developing rapidly, including numerous exchanges, e-wallets, miners, and stablecoin issuers.

Nevertheless, Japan's economic growth has been damage by the bad debts of financial institutions. When the bank's problem with non-performing assets started to show up, the Japanese government changed its approach a lot, especially before and after the Financial Regeneration Law was passed, to help financial institutions with weak economic structures get back on their feet. After the Financial Regeneration Act goes into effect, financial institutions with lousy management will be able to leave the market in a controlled way, and the non-performing debts of bankrupt financial institutions will be given to sorting and recycling institutions to clean up and recover over time. 

Using early corrective measures and the Act on the Early Improvement of Financial Functions, the Japanese government has strengthened the capital structure of good financial institutions. It took 16 years for the Japanese banking crisis to get better, and there are many things to learn from it.

\section{What are Non-Performing Assets}
A Non-performing Asset (NPA) is a designation for delinquent loans or advances. Debt is in arrears when principal or interest payments are late or missed. A loan defaults when the lender deems the loan arrangement to have been breached, and the debtor cannot satisfy his financial commitments. 
NPAs can be categorised as substandard, dubious, or loss assets based on the length of time overdue and the likelihood of repayment.

Therefore, NPAs impose a financial burden on the lender; a large number of NPAs over time may signal to regulators that the bank's financial health is in peril. 

\subsection{Sub-Classifications for Non-Performing Assets (NPAs)}
Usually, a grace period is given by lenders before designating an asset as non-performing. The lender or bank will then classify the NPA into one of the subsequent sub-categories: 

1. Typical Assets 
They are NPAs with a typical risk level that has been past due for 90 days to 12 months. 

2. Non-standard Assets: 
 Non-standard assets have more than a year's worth of past-due balances. They have a significantly increased risk level in addition to a borrower with substandard credit. Because they are less confident that the borrower will ultimately repay the total amount, banks typically assign haircuts (reductions in market value) to such NPAs.

3. Doubtful Debts
In the category of "doubtful debts," non-performing assets have been past due for at least 18 months. Most of the time, banks doubt that the borrower will ever pay back the entire loan. This type of NPA significantly affects how risky the bank is as a whole. 

4. Loss Assets
These are non-performing assets that haven't been paid back for long. With this class, banks are forced to accept that the loan will never be paid back and must record a loss on their balance sheet. The whole amount of the loan must be entirely written off.

\subsection{How NPAs Work}

First, non-performing assets are listed on a bank's or other financial institution's balance sheet. After a long time of not paying, the lender will force the borrower to sell any assets that were put up as collateral for the loan. If no assets were put up as collateral, the lender could write off the asset as a bad debt and sell it to a collection agency at a discount. 

Most of the time, debt is considered "non-performing" if loan payments haven't been made for 90 days. The average amount of time is 90 days, but it can be shorter or longer depending on the terms and conditions of each loan. A loan is an example of a non-performing asset.

For example, let's say a company with a $10 million loan and monthly payments of $50,000 just for the interest doesn't pay for three months in a row. The lender may label the loan as not being paid back to meet regulatory requirements. A loan is also considered a non-performing asset if a company pays all the interest but can't pay back the principal when it comes due. 

When non-performing assets are listed on the balance sheet, it stresses the lender a lot. When interest or principal isn't paid, the lender's cash flow goes down, which can mess up budgets and lower income. Loan loss provisions, which are money set aside to cover possible losses, reduce the amount of money that can be used to lend to other people in the future.

\FloatBarrier
\begin{table}[H]
\begin{tableminipage}{\textwidth}
\caption{Dimensions of the specimens}
\begin{tabularx}{\linewidth}{ 
   >{\centering\arraybackslash}X
   >{\centering\arraybackslash}X
   >{\centering\arraybackslash}X
   >{\centering\arraybackslash}X
   >{\centering\arraybackslash}X
   >{\centering\arraybackslash}X
   >{\centering\arraybackslash}X}
\hline
{\bf LAST} & {\bf PREVIOUS} & {\bf MIN} & {\bf MAX} & {\bf UNIT} & {\bf FREQUENCY} & {\bf RANGE} \\
\hline
1.2 & 1.2 & 1.1 & 8.4 & \% & semiannually & Mar 1999 - Sep 2021 \\

\hline
\end{tabularx}
\end{tableminipage}
\label{table1}
\end{table}
\FloatBarrier

\section{Japan's Npas History and Present}

The non-performing loan harmed Japan's financial market but improved the nation's banking system. In Japan, the financial crisis began in 1994. The Japanese government issued the "Resolution and Collection Corporation, RCC" to prevent the spread of the financial crisis. This company is one of Japan's methods for dealing with non-profit financial organisations. It was also utilised by the United States to mitigate the financial crisis of the early 1990s.

The Japanese government passed the Bank Deposit Insurance Act after the RCC ran. The deposit insurance company is accountable for directing the operation of the institution. The amount of capital is 212 billion yen. The RCC is able to acquire NPAs from both the closed and lively financial institutions. 

In addition, the Deposit Insurance Organisation established a "Purchase price reviewing department" to ensure that the price of NPAs is acceptable and reasonable.It is composed of lawyers, accountants and real estate appraisal experts to set relevant pricing procedures and price-setting mechanisms for non-performing debts of different natures to avoid losses caused by improper pricing.

As shown below, Japan's Non Performing Loans Ratio from Mar 1999 to Sep 2021. In July 2016, Japan's ratio of nonperforming loans was 1.5, steadily declining to 1.1 by January 2018. From 2018 to 2020, the ratio remained constant at 1.1. In January 2021, meanwhile, due to the global spread of COVID-19 and other unforeseen circumstances, the ratio of NPAs increased to 1.2. 

\FloatBarrier
\begin{figure}[H]
\includegraphics[width=\linewidth]{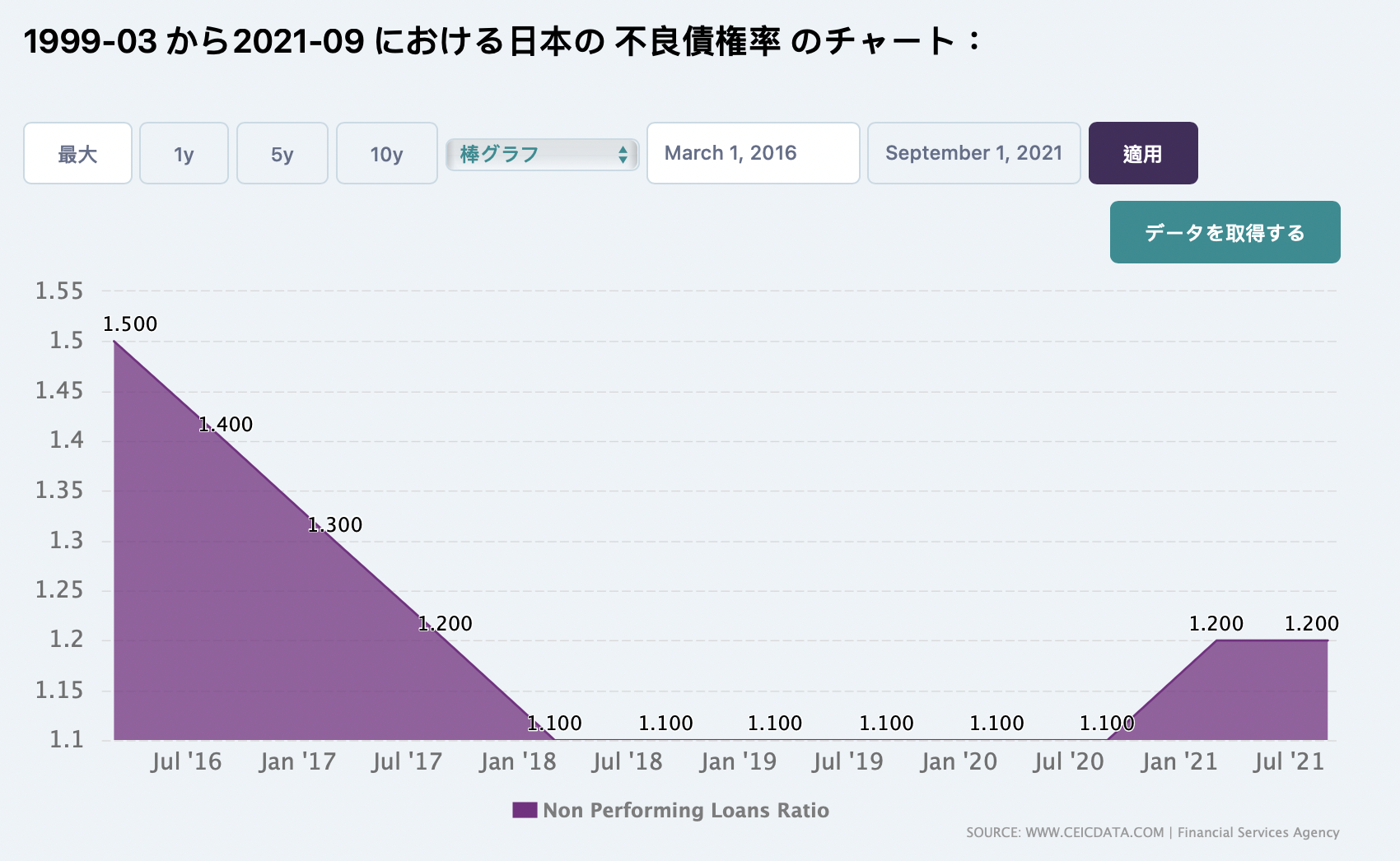}
\caption{Japan's Non Performing Loans Ratio from Mar 1999 to Sep 2021.}
\label{Figure1}
\end{figure}
\FloatBarrier

In the following chart, we can see that the latest ratio in September 2021 is 1.2.

\section{Impact of NPAs in Japan}
In the preceding section, we analysed the historical and current NPA situations of Japan. In this section, we will examine how 

\vspace{100pt}
\noindent "the problem of non-performing assets," the situation in which banks are saddled with a large number of non-performing assets that do not decrease and remain at a high level for an extended period of time, contributes to the Japanese economy's prolonged stagnation. 

The problem of non-performing loans has the following negative effects on the economy: 

1) The erosion of banks' profitability;

2) Stagnation of economy; and 

3) Cautious behaviour of corporations and consumers 

\subsection{The Erosion of Banks' Profitability}

In the case of a lost asset, the loss has been identified, and the amount is not deducted. The NPAs reduce the profitability of banks as a result of a rise in operating expenses and a decline in their interest margins. According to studies, banks with a high proportion of nonperforming assets incur "carrying costs" that reduce their profitability. 

The greater the operating costs of a bank, the lower its cost efficiency and consequently its profits. Included in operating expenses are employee wages and salaries as well as costs associated with branch offices. These costs have a negative effect on bank profitability. A rise in nonperforming assets is likely to have a negative impact on the profitability of banks due to the large amount of provisioning requirements, which act as a drain on the profitability of banks.

Consequently, the provisioning and carrying costs of NPAs act as drain on the profitability of the banks. 

\subsection{Stagnation of Economy}

Economic Stagnation is characterised by slow or flat growth. If annual real economic growth is less than 2 percent, periods of high unemployment and involuntary part-time employment are prevalent. Moreover, Stagnation can occur as a temporary condition, such as a growth recession or temporary economic shock, or as a result of the economy's long-term structural condition.
 
It frequently involves high unemployment and underemployment, as well as an economy that performs generally below its potential. Various economic and social factors can cause lengthy or brief periods of Stagnation.
 
High NPA ratios and restrictions on corporate business have a direct effect on the lending attitudes of banks. Since banks can assume that companies and financial institutions will not be able to repay in the future, they will not lend any money to them. Therefore, banks have become reluctant to grant loans and accept zero-percent risks. The creation of new credit is prohibited. 
 
Nevertheless, the majority of small and medium-sized businesses lack effective alternatives to internal funding and bank borrowing. If banks do not lend them money, the businesses will fail. It results in fewer employment opportunities for the general public. Assuming that the labour force will remain unchanged and other factors will not change, the unemployment rate will rise, and economic growth will slow. It causes economic Stagnation.

\subsection{Cautious Behaviour of Corporations and Consumers}

As non-performing loans increase, the bank's image suffers, and the public loses faith in banks. The depositors may withdraw their funds, causing banks to experience liquidity issues. In addition, the higher NPAs ratio results in corporations and consumers being more cautious. (NPAs ratio is the ratio of a bank's nonperforming assets, or bad loans, to its total assets.) 
The lack of liquidity prevents banks from lending to other economically productive activities. Investment restraint may slow the economy, resulting in unemployment, inflation, bear market, et cetera.

For the banks, in order to maintain their profit margins, banks will be compelled to raise interest rates, further harming the economy. Due to the high cost of interest, increasing interest rates can significantly reduce the number of people who are willing to purchase. 

Thus, as a result of declining confidence in the financial system, corporations and consumers exhibit cautious behaviour.

\section{Comparative Study Between Banks Having Less and High NPAs}

The non-performing assets (NPAs) point to difficulties for banking and financial institutions, which is not unexpected. The failure of a great number of them has created an emergency situation for the customers and the investors. A gross non-performing asset ratio that is extremely high is an indicator that the bank's assets are in a precarious state. A low gross nonperforming asset ratio, on the other hand, indicates that the bank's assets are in healthy condition. 

In point of fact, no bank wants to have a non-performing asset (NPA) on its books, but when the economic cycle gets worse, it is sometimes unavoidable. Therefore, a bank is said to have good management if its ratio of non-performing loans is less than 1 percent. It is continuously maintained by private banks like the Housing Development Finance Corporation (HDFC), among others.

\section{What are Crypto-currency}

Cryptocurrency is a peer-to-peer digital exchange system that generates and distributes units of cash using cryptography. Without a central authority, this system requires decentralised transaction verification. Transaction verification validates transaction amounts and the payer's ownership of the money they seek to spend and prevents double payment in currency units. This form of verification is known as "mining." 

Mining also generates wealth in the form of new currency units. It calls for robust, secure mining algorithms, which are subject to public scrutiny, and to keep them decentralised. Cryptocurrencies utilise various mining technologies in accordance with their specific requirements. Different cryptocurrencies, for instance, prioritise limiting the number of transactions validated per unit of time, while others prioritise establishing quick, lightweight services. Some mining techniques are purposefully memory intensive, while others are computationally costly. 

This study examines the following cryptocurrencies: Bitcoin (BTC) and Ripple (XRP) in Japan. These two cryptocurrencies are the most intriguing, extensively used, and have the highest capitalisation and transaction rates. In addition, they highlight the primary mining algorithms. 

\subsection{Bitcoin vs. XRP}

From the outside, the cryptocurrency investment universe appears to be limited to bitcoin. Bitcoin is the most popular cryptocurrency regarding market capitalization and investor interest. Those interested in portfolio diversification and experimenting with coins that offer a novel take on the concept of digital currencies have additional options. Ripple (XRP) is one of them. In May 2022, the cryptocurrency ranked sixth in total market capitalization. Consider how XRP differs from Bitcoin and other leading digital currencies. 

\subsection{Bitcoin}

Bitcoin (BTC) is a cryptocurrency, a digital asset meant to operate as a medium of exchange and a form of payment independent of a central bank or single administration, hence eliminating the need for third parties in financial transactions. It is paid to blockchain miners for their efforts in verifying transactions, and it can be acquired on a number of exchanges. In 2009, an anonymous developer or group of unknown developers using the name Satoshi Nakamoto released it to the world.
 
Bitcoin relies on a public blockchain record that supports a digital currency used to make purchases of goods and services. The bitcoin network is founded on the blockchain idea, a public ledger of verifiable transactions and record keeping. 

Continuously, miners validate transactions and add them to the Bitcoin blockchain. Upon successfully verifying transactions, miners are rewarded with BTC in return for their time and the computational power required to validate the ledger in this circumstance.
 
It has become the most well-known cryptocurrency worldwide. Its prominence has encouraged the creation of several more cryptocurrencies. These rivals either want to supplant it as a payment mechanism or employ utility or security tokens in other blockchains and new financial technologies.

\subsection{XRP}

Ripple is a blockchain-based digital payment network and protocol with its cryptocurrency, XRP. Ripple's primary function is a payment settlement asset exchange and remittance system, similar to the SWIFT system for international money and security transfers used by banks and financial intermediaries dealing in several currencies.
 
The cryptocurrency's token was pre-mined and represented by the ticker sign XRP. Ripple is the corporation and network name, whereas XRP is the cryptocurrency coin. XRP is intended to function as an intermediary mechanism of exchange between two currencies or networks – a form of interim settlement layer currency. Chris Larsen and Jed McCaleb co-founded Ripple, which was introduced in 2012 and co-founded by Larsen and McCaleb.
 
It operates on an open-source and decentralised peer-to-peer platform that enables the transmission of money in any form, including dollars, yen, euros, and cryptocurrencies such as litecoin. Ripple is a worldwide payments network whose clients include large banks and financial industry firms. In its products, XRP is used to ease the rapid conversion of other currencies.

\section{Risks and Returns of Trading Cryptocurrency}

Cryptocurrencies are the potential to reshape the financial world as we know it and put into question the entire existence of conventional economic infrastructure. What are the advantages and disadvantages of cryptocurrencies?

\subsection{Risks of Trading Cryptocurrency}

The risks associated with trading cryptocurrencies include their volatility, lack of regulation, susceptibility to error and hacking, etc.

1) Volatility 

Unexpected changes in market presumption can lead to sharp and sudden price movements. It is prevalent for the value of cryptocurrencies to plummet by hundreds or even thousands of dollars in a short period of time. As quickly as the price of a cryptocurrency can soar to dizzying heights, it can also plummet to terrifying depths. 

Therefore, this may not be investors best option for seeking stable returns. Fundamentally, the cryptocurrency market thrives on speculation, and its relatively small size makes it more susceptible to price fluctuations. This can wreak havoc on the value of coins, which is one of the primary disadvantages of cryptocurrencies.

2) Not considered as a long-term investment

It is critical to remember that cryptocurrencies have existed for less than a decade, despite their widespread recognition and continued growth in popularity. The concept did not emerge until 2008 when a white paper on Bitcoin was published. In contrast, stock markets can look back on centuries of history. For example, the London Stock Exchange was established in 1801. Gold has been a dependable store of value for aeons. But cryptocurrencies? As an investor, you must have the courage to enter these uncharted waters because nobody knows what the future holds for cryptocurrencies.

3) Vulnerable to security risks

Cryptocurrencies may not have the risks associated with relying on centralised intermediaries, but that does not mean they are completely secure. Technical errors, human errors, and hackers cannot be prevented with absolute certainty. As a cryptocurrency owner, you risk losing the private key that grants you access to your coins along with your entire holdings. And then there is hacking, phishing, and all the other malicious attempts to gain control. This is something that seasoned investors watch out for, but novice investors are more likely to fall victim to such traps.

4) Serious scalability issues

The trading of cryptocurrencies includes significant risks, including hard forks and discontinuance. Before trading these items, you should familiarize yourself with their dangers. When a hard fork happens, there may be significant price volatility surrounding the event, and we may cease trading if the underlying market's prices are not reliable. You could be forgiven for believing that digital currencies operate at the speed of light, and to a certain extent, they do. However, they encounter significant obstacles at a particular scale that make their widespread implementation difficult. 

The blockchain has reached "certain capacity limitations" that slow the rate at which transactions can be processed, as acknowledged by the developers of Ethereum. This can be a frustrating experience for transaction participants, not to mention that it may result in financial losses.

\subsection{Returns of Trading Cryptocurrency}

Investments inevitably include both risks and profits. How much risk you take will determine how much money you gain. Despite the fact that cryptocurrencies are a relatively recent creation (Bitcoin, for instance, was founded in 2009), they are here to stay, with all their advantages. If you know how to access it, the world of cryptocurrencies has a great deal to offer, from the possibility of substantial profits to ultra-secure, transparent infrastructure available 24/7. 

1) High risk, but a high possibility for a high return 

There are already over 10,000 cryptocurrencies available on the market, and each has its features. However, all cryptocurrencies share specific characteristics, such as the propensity to experience abrupt price increases and decreases. The primary drivers of prices are the supply of coins from miners and the demand for them from buyers. These supply-demand dynamics can yield substantial profits. The price of XRP, for instance, roughly doubled between September 2022 and October 2022, yielding a considerable profit for early investors.

2) Inherently secure

Some of the greatest benefits of cryptocurrencies are not related to the currencies themselves but to the infrastructure supporting them. The blockchain is a decentralised data storage ledger that records every transaction executed. A block cannot be removed from the blockchain after it has been added. And because the blockchain is distributed across multiple computers, it is difficult for a hacker to access the entire chain simultaneously. Therefore any data saved in it is safe forever.

3) Fairer, more transparent financial system

Compared with traditional banks, the cryptocurrencies financial system mainly depends on third-party intermediaries that process transactions. This means that if you conduct a transaction, you are placing your trust in one or more of these intermediaries, which the recession of the early 2000s caused many people to question. Blockchain technology and cryptocurrencies provide an alternative. They can be viewed by anyone, allowing you to participate in financial markets and conduct transactions without intermediaries.

4) Cryptocurrency is constantly exchanged. 

Another benefit of cryptocurrencies over banks is that crypto markets are always accessible. You do not need to wait until the NYSE, NASDAQ, or any other exchange to begin trading for the day to buy, sell, or trade cryptocurrencies, as coins are mined, and transactions are recorded around the clock. This has had such an impact that traditional stock exchanges are examining the possibility of trading stocks outside of regular banking hours, although this may be some time away. Therefore, crypto may be the best way for investors constantly on the move to generate returns outside regular business hours.

5) Cryptocurrency is superior to inflation

Since cryptocurrencies are not tied to a single currency or economy, their value reflects global demand rather than, for example, national inflation. However, what about the inflation of cryptocurrencies? As an investor, you can, for the most part, rest easy. As a result, there is no inflation as the number of coins is capped and cannot spiral out of control. Some cryptocurrencies such as Bitcoin have an overall cap, while others such as Ethereum have an annual cap. Regardless, this method keeps inflation at bay.

\section{Crypto-currency in Japan}

In Japan, cryptocurrencies are legal. The Payment Services Act defines "crypto-assets" as non-fiat payment methods that can be used for unspecified purposes. There are no restrictions on cryptocurrency ownership and investment. In Japan, cryptocurrencies are legal. The Payment Services Act defines "crypto-assets" as non-fiat payment methods that can be used for unspecified purposes. There are no restrictions on cryptocurrency ownership and investment.
 
In addition, Japan has anti-money laundering (AML) cryptographic regulations. The Act on the Prevention of the Transfer of Criminal Proceeds outlines the anti-money-laundering regime applicable to exchange providers in Japan. The Guidelines for Anti-Money Laundering and Combating the Financing of Terrorism, which have been in effect since February 19, 2021, further specify this framework.

The Payment Services Act identifies exchange providers as businesses that offer any of the following crypto-asset exchange services: 
Sale, purchase, and exchange of crypto assets;
Inter-mediation, brokering, or acting as an agent in 1;
Management of customer funds in connection with 1 or 2;
Management of crypto assets for the benefit of a third party.

\section{XRP is More Popular in Japan Than the U.s.}

\begin{figure}
\includegraphics[width=\linewidth]{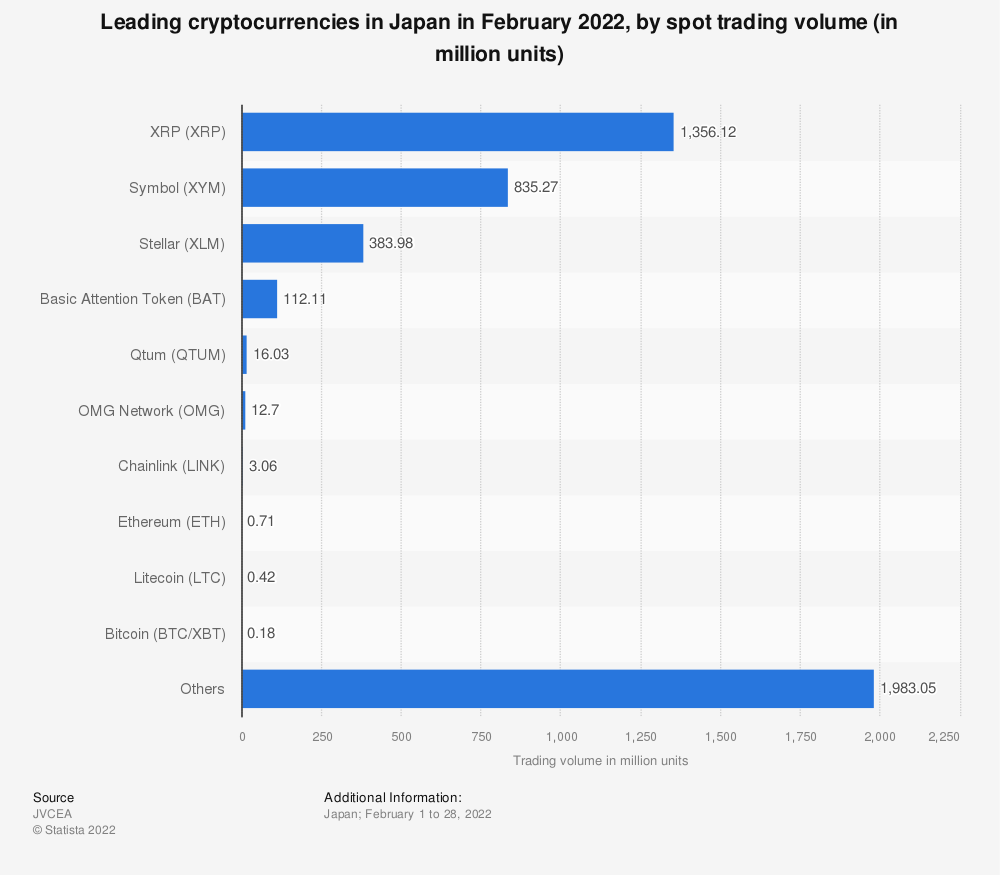}
\caption{Leading cryptocurrencies in Japan in February 2022, by spot trading volume (in million units).}
\label{Figure2}
\end{figure}

With nearly 1.36 billion units traded in February 2022, XRP was the cryptocurrency with the highest spot trading volume in Japan. In the same month, Bitcoin, which had the highest trading value, had a trading volume of about 180 thousand units as shown in \hyperref[Figure2]{Figure 2}.

\FloatBarrier
\FloatBarrier

With a lawsuit from the Securities and Exchange Commission, Ripple has had a turbulent year in the US over the past year. However, Japan is where Ripple is most well-known. In actuality, a significant portion of XRP's price movement has been caused by Japanese investors buying the cryptocurrency and driving the price up. 

\subsection{Why Japanese like XRP}

Many Westerners don't understand why Ripple has become so popular in Japan. Once one is familiar with Japanese customs and the confidence that Japan's largest venture capital fund has in Ripple, the whole thing becomes clear. 

The partnership between the most reputable venture capital firm in Japan, SBI Holdings, and Ripple has been the primary driver of Ripple's success in Japan. In 2017, SBI Holdings invested close to 300 million in Ripple. In the SEC lawsuit, this investment was cited as an improper sale of a security to SBI Holdings. Even the founder of SBI Holdings and a board member of Ripple tweeted about the new all-time high of XRP, which was also referenced in the lawsuit. 

Despite Ether's position as the second largest cryptocurrency by market cap, some Japanese crypto investors may find XRP to be a more appealing investment. 

According to the results of a survey published on Twitter by Japanese cryptocurrency exchange BITMAX on June 12, Bitcoin's popularity among Japanese traders is comparable to that of the XRP token's. 25\% of the 1,498 respondents said XRP was their favourite cryptocurrency, compared to 26\% who said BTC was their favourite. And 9\%, Ethereum ETH ranked third.

\FloatBarrier
\begin{figure}
\includegraphics[width=\linewidth]{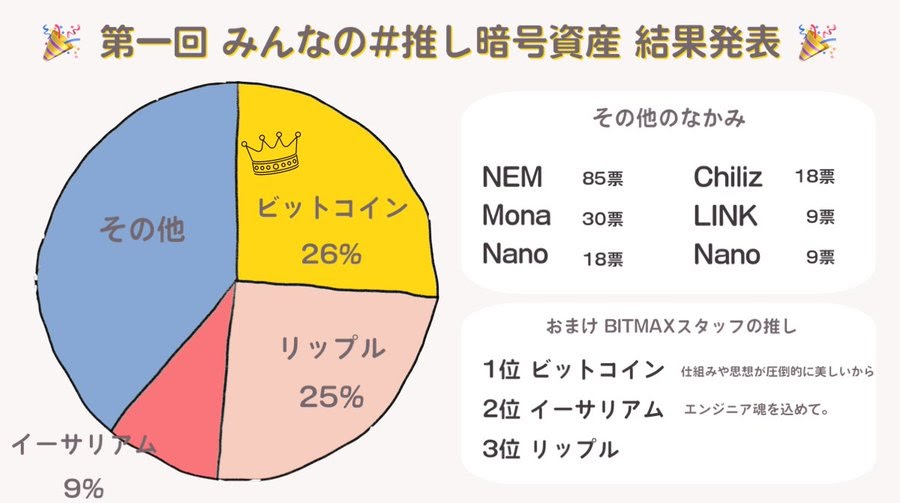}
\caption{What is your preferred coin? BITMAX survey. Gold = BTC; Red = ETH; Pink = XRP; Blue = Others}
\label{Figure3}
\end{figure}
\FloatBarrier

It is important to note that a small sample size, such as that of the aforementioned survey, may not necessarily render the results conclusive. 

In December 2019, however, the Japan Virtual Currency Exchange Association (JVCEA) reported similar findings. XRP was second in terms of yen-denominated digital asset holdings, behind BTC, with ETH in third place.

\section{Crypto Impact the Stock Market in Japan}

Upon factoring in cryptocurrency's volatility, there is still some correlation between cryptocurrency and stock prices. Price movements in cryptocurrencies are susceptible to many of the same market forces that impact stock prices. Cryptocurrency prices follow the same general patterns as stock prices because investors and traders treat them similarly.
 
\subsection{What Factors Affect Stock and Cryptocurrency Prices}

Investors have traditionally favoured the stock market. Stock market performance and price movements are highly scrutinised, so their causes have been extensively researched.
 
In the table below, we've included some variables that can impact the value of stocks and cryptocurrencies.

\FloatBarrier
\begin{figure}[H]
\includegraphics[width=\linewidth]{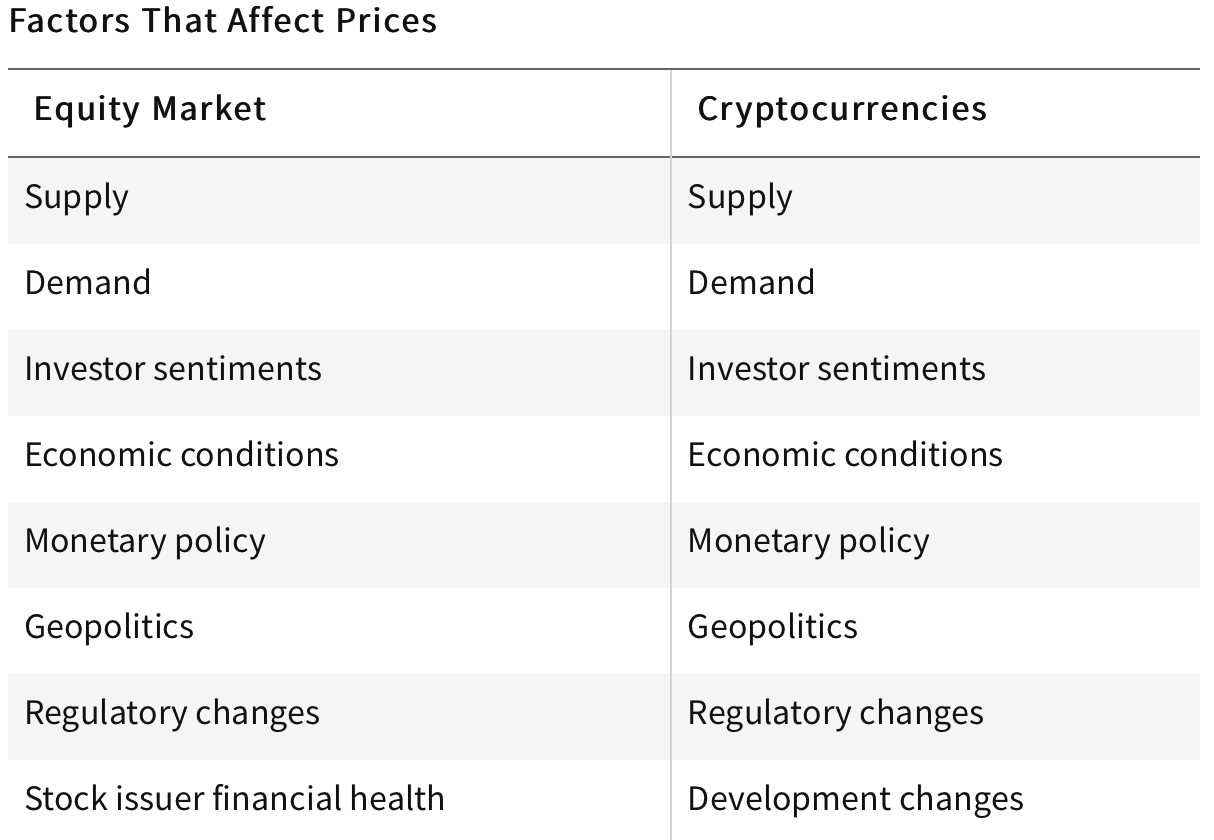}
\caption{Factors impact the value of stocks and cryptocurrencies}
\label{Figure4}
\end{figure}
\FloatBarrier

\subsection{1. Supply and Demand}
Cryptocurrencies like Bitcoin have a finite supply. Contrarily, the total amount of bitcoin that may be mined or issued in the future is not capped in any way with other cryptocurrencies. However, the number of outstanding shares of a company's stock may be managed and is ultimately supported by the success of the business that issued the stock. 

It's commonly accepted that market supply and demand play a significant role in setting prices. These factors also affect the value of stocks, and it appears that Bitcoin is also. There will only ever be a maximum of 21 million Bitcoin, so as demand grows, does the price. To capitalize on the trend, investors are looking into other cryptocurrencies. 

\subsection{2. Investor Beliefs and Expectations}
Investor sentiment is one of the most critical variables that affect prices. Investor sentiment in the equity market refers to investors' expectations for the market. They are divided into two groups based on this belief: those who think prices will rise and those who think prices will fall. Then, based on their outlook, they make investment decisions. 

\subsection{3. Regulation}
Securities and other regulators in the nation of origin often do extensive research on equities, often known as stocks. Further, stock exchanges monitor companies whose shares are traded there and may even delist them if they find companies against government polices to do so. In no way is this a guarantee, but it is more excellent protection than you would have with any other cryptocurrency investment.
 
The decentralized nature of cryptocurrency networks enables direct monetary transactions between users without a central authority. In contrast to the transparency of traditional stock trades, the anonymity of cryptocurrency transactions is a significant selling point for many investors.

Two years after the original publication of the FATF's Guidance for a Risk-Based Approach to Virtual Assets (VAs) and Virtual Asset Service Providers (VASPs)" in March 2021, the FSA sent a Request for Notification of Originator and Beneficiary Information upon Crypto Assets Transfer ("Request") to the JVCEA. The Request said that by April 2022, JVCEA should have introduced self-regulatory norms concerning the travel rule. With the advent of April 2022, JVCEA included the Travel Rule in the Association's self-regulatory guidelines for transferring crypto assets.

\section{Cryptocurrency Prices vs. Stock Prices}
To begin with, it's important to remember that one essential ingredient needs to be added to every market: buyers and sellers. Large institutional funds and individual investors contribute to the stock market's strong liquidity. Retail traders already engaged in the stock market are also likely to participate in cryptocurrencies. 

The PDT restrictions that require a minimum of \$25,000 in equity to day trade are not present in the cryptocurrency market, making access considerably easier. Cryptocurrencies appeal to new traders because there are no admission requirements, no PDT regulations, and the market is open around the clock, seven days a week. 

This is especially true among the younger generations, who comprise most millennial and Gen Z populations, and vice versa. An increase in the number of market participants may cause a similar increase in the number of other market participants. 

\FloatBarrier
\begin{figure}[H]
\includegraphics[width=\linewidth]{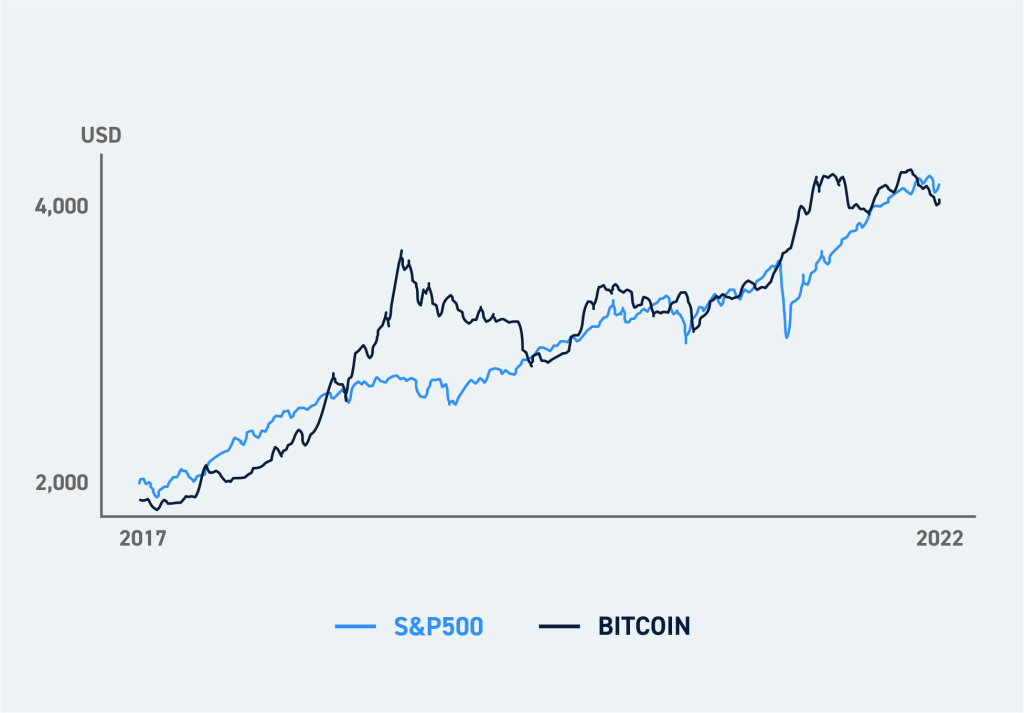}
\caption{S\&P 500 index and Bitcoin trends.}
\label{Figure5}
\end{figure}
\FloatBarrier

The crypto market does show some links with the stock market. The stock prices of companies with significant investments or operations in cryptocurrency tend to track the underlying price movement of such assets. Contrary to popular belief, however, stock market indices still need a clear and consistent relationship with the cryptocurrency markets. In the same way that the S\&P 500 index typically determines the stock market's direction, Bitcoin price movement tends to indicate the general trend of cryptocurrencies. 

Thus, there is a correlation between the popularity of crypto markets and the popularity of specific financial needs, industries, and even topics. Largely, the cryptocurrency markets are strongly associated with particular stock markets. For instance, firms specializing in cryptocurrency mining and blockchain technology. Bitcoin exchanges will be affected by the market's broader trends and price fluctuations. For instance, if bitcoin is experiencing a breakout, the shares of a firm involved in bitcoin mining can also experience a breakout. 

\section{Transaction Cost in Stock Market and Crypto Market}
Costs associated with purchasing and selling products or services are referred to as "transaction costs." This concept has given rise to whole businesses whose sole purpose is facilitating trade. In finance, "transaction costs" refers to the fees paid to brokers through commissions and spreads. It is also defined as the difference between the price a dealer pays for security and the price that the buyer pays. 
Initially disregarded by traders and investors as a niche sector, cryptocurrencies have become a popular asset class. If you wish to participate, you must utilize a cryptocurrency exchange to acquire exposure to this expanding industry. Trading cryptocurrencies is comparable to trading on a stock exchange. You may get charged via trade on cryptocurrency exchanges. Most cryptocurrency exchanges use a tiered system that charges a portion of your 30-day trading volume to determine prices. Transferring money to and from your bank account may include wire costs, mining fees, account fees, spot fees, tiered transaction fees, etc. 

When an investor buys or sells shares of stock, the price may consist of two components: the cost of the shares plus any transaction fees levied by the brokerage company. This cost is known as the commission. Recently, online brokers have been engaged in an all-out pricing war. Many of the largest online brokers provide fee-free stock trades, but most charge a cost for trading mutual funds. For stock transactions, most full-service brokers charge between 1\% and 2\% of the entire purchase price, a flat fee, or a mix of both. They provide investors with financial planning and investment advice and execute transactions on their behalf.
 
Therefore, the market for cryptocurrencies and the market for stocks both have transaction costs, but these costs are indicated in different regions of the markets. In the stock market, it is customary to pay a commission. On the other hand, there is no commission charge for broker in the cryptocurrency market. You don't need to locate a broker to assist you in purchasing and selling cryptocurrency, which means that you can do it all by yourself. 

\section{Conclusion}
In this paper, we look at some examples of non-performing assets in Japan and discuss cryptocurrency.  

Non-performing assets (NPAs) point to difficulties for banking and financial institutions, which is not unexpected. In July 2016, Japan's ratio of nonperforming loans was 1.5, steadily declining to 1.1 by January 2018. In January 2021, due to the global spread of COVID-19 and other unforeseen circumstances, the ratio of NPAs increased to  1.2. The ratio for the period from Mar 1999 to Sep 2021 is shown in section "Japan's NPAs history and present". A bank is said to have good management if its ratio of non-performing loans is less than 1 percent. It is continuously maintained by private banks like the Housing Development Finance Corporation (HDFC), among others.

The similarity between cryptocurrency and equities prices might signal that cryptocurrency values follow equity price patterns. As an investment tool in the crypto market, the Japanese would favor XRP. There are hazards regardless of whether we invest in the stock or cryptocurrency market. The stock market has greater oversight than the cryptocurrency sector. The stock market has fixed trading hours and is centralized, whereas the cryptocurrency market trades around the clock and is decentralized. 

Cryptocurrencies make international expansion simple for investors. The alternative, crypto-trading, may be helpful for global investors, given that investing in foreign stocks is time-consuming. Before a transaction is verified, cryptocurrency is not subject to regulatory organizations such as Central Institutions or other banks. This is one of the key reasons for its organic evolution. 

Due to its decentralized network structure, which utilizes a peer-to-peer payment network based on blockchain technology, virtual currencies provide a great alternative to the current monetary system. The financial services business can be revolutionized by blockchain technology, particularly in terms of automating market monitoring and post-trade event processing. 

To realize the total potential value of restricted mutual public blockchains and smart contracts, however, would require substantial revisions to business practices and investments from enterprises, particularly on the buy- and sell-sides of the industry. So, the market should monitor possible legal changes. Strong governance is essential for blockchain to safeguard participants, investors, and other stakeholders. It ensures the system's ability in the face of systemic risk, privacy problems, and cyber-security attacks.

\nocite{*}
\bibliography{bibliography}

\end{document}